\documentclass[aps,prb,onecolumn,showpacs,floatfix,superscriptaddress,amssymb,amsmath]{revtex4}
\usepackage{graphicx}
\usepackage{dcolumn}
\usepackage{bm}
\usepackage{psfrag}

\newcommand{\be}{\begin{equation}}
\newcommand{\ee}{\end{equation}}
\newcommand{\bea}{\begin{eqnarray}}
\newcommand{\eea}{\end{eqnarray}}

\newcommand{\ba}{\begin{array}}
\newcommand{\ea}{\end{array}}

\begin{document}

\title{Quantum transport in a curved one-dimensional quantum wire with
spin-orbit interactions}

\author{Erhu Zhang, Shengli Zhang, Qi Wang}
\affiliation{Department of Applied Physics, Xi'an Jiaotong
University, \\Xi'an 710049, P. R. China}

\date{\today}

\begin{abstract}
The one-dimensional effective Hamiltonian for a planar curvilinear
quantum wire with arbitrary shape is proposed in the presence of the
Rashba spin-orbit interaction. Single electron propagation through a
device of two straight lines conjugated with an arc has been
investigated and the analytic expressions of the reflection and
transmission probabilities have been derived. The effects of the
device geometry and the spin-orbit coupling strength $\alpha $ on
the reflection and transmission probabilities and the conductance
are investigated in the case of spin polarized electron incidence.
We find that no spin-flip exists in the reflection of the first
junction. The reflection probabilities are mainly influenced by the
arc angle and the radius, while the transmission probabilities are
affected by both spin-orbit coupling and the device geometry. The
probabilities and the conductance take the general behavior of
oscillation versus the device geometry parameters and $\alpha $.
Especially the electron transportation varies periodically versus
the arc angle $\theta _{w}$. We also investigate the relationship
between the conductance and the electron energy, and find that
electron resonant transmission occurs for certain energy. Finally,
the electron transmission for the incoming electron with arbitrary
state is considered. For the outgoing electron, the polarization
ratio is obtained and the effects of the incoming electron state are
discussed. We find that the outgoing electron state can be spin
polarization and reveal the polarized conditions.
\end{abstract}

 \keywords{}

\maketitle

\section{Introduction}

Recently, a new subdiscipline of condensed matter physics,
spintronics, is emerging rapidly and generating great
interests$^{[1,2]}$. Much attention has been focused, especially, on
the spin-dependent transport dominated by the spin-orbit interaction
(SOI)$^{[3,4]}$ in the low dimensional semiconductor nanostructures,
such as wires, rings, spirals, and other structures.\ For example,
the influence of spin on electrons moving in a mesoscopic ring have
been studied in several contexts$^{[5-10]}$ since the
original proposal of the spin field effect transistor by Datta and Das$%
^{[11]}$.

On the other hand, many low-dimensional systems are curvilinear and
more complicated in geometry, such as the V-shaped quantum
wire$^{[12]}$, the spiral inductors$^{[13]}$, nanotubes$^{[14]}$,
and so on. Due to the potential of providing new physical features
and new functionalities for electronic devices$^{[15]}$, the effects
of the curvilinear geometry on electronic states and electron
transports are the subjects of many recent works$^{[16-20]}$.
Especially, the influence of geometry on the SOI is considered by
some groups$^{[21,22]}$. The systems concerned are
quasi-one-dimensional circles or two-dimensional (2D)
nanostructures. However, for one dimensional (1D) system, the
electron behavior in curved quantum wires with arbitrary shape is
still a challenging question. The purpose of this paper is to study
the curvature effects on the electron transport through a 1D\
curved\ nanostructure taking into account the SOI.

This paper is organized as follows. In Sec. II, we investigate a 1D
planar curvilinear quantum wire with arbitrary shape. The effective
Hamiltonian is derived using perturbation theory. In Sec. III we
study the one-electron energy spectrum and eigenstates of
curvilinear quantum wire. Then we consider the corresponding
transmission of electrons. A detailed numerical discussion is given.
Conclusions and remarks follow in Sec. IV.

\section{EFFECTIVE HAMILTONIAN FOR\ A 1D PLANAR CURVILINEAR QUANTUM WIRE}

We consider the adiabatic motion of electrons bounded on a planar
quantum
wire as shown in Fig. 1(a). Under the effective-mass approximation, the Schr%
\"{o}dinger equation for an electron in the presence of Rashba
SOI$^{[3]} $ is given by

\begin{align}
\lbrack \frac{{\bf p}^{2}}{2m}+V({\bf r})+\hat{H}_{SO}]\Psi & =E\Psi
{\bf ,}
\\
\hat{H}_{SO}=\alpha {\bf \hat{\sigma}}\cdot {\bf n}\times {\bf p},&
\nonumber
\end{align}%
where $m$ is the effective electron mass, $\hat{\sigma}$ is the
Pauli matrices and ${\bf n}$ is the surface normal. Here $V({\bf
r})$ represents the potential which confines electrons to the wire,
while the Rashba term results from the asymmetric confinement along
the direction perpendicular to the plane. The parameter $\alpha $ is
the Rashba parameter which represents the average electric field
along the ${\bf n}$ direction.

In order to obtain the effective Hamiltonian for electrons on a 1D
wire, we take the same approach as da Costa's$^{[23]}$. Let ${\bf
a}(s)$ parametrically specify the 1D planar curve where $s$ is the
arc length along the curve. Any point in the surface surrounding the
curve can be expressed as follows:

\begin{equation}
{\bf r}(s,u)={\bf a}(s)+u{\bf b}(s),\text{ }\left| u\right| \leq \frac{%
\varepsilon }{2},
\end{equation}%
where ${\bf b}(s)=\frac{1}{\kappa }\frac{d^{2}{\bf a}(s)}{ds^{2}}$
is the normal of ${\bf a}(s)$ and $\kappa =\left| d^{2}{\bf
a}(s)/ds^{2}\right| $ is the curvature of the curve, the curve width
$\varepsilon $ is small and is assumed to be less than the curvature
radii. The twosome $\{s,u\}$\ forms a curvilinear coordinate system
in the 2D plane. The concerned electron moving along the wire is
confined near the curve ${\bf a}(s)$ in the normal direction by a
potential $V(u)$.

In this curvilinear coordinate system, the corresponding metric
tensor is

\begin{equation}
G_{ij}=\left(
\begin{array}{cc}
(1-u\kappa )^{2} & 0 \\
0 & 1%
\end{array}%
\right) .
\end{equation}%
So the wave function should be normalized according to the condition

\begin{equation}
\int \left| \Psi \right| ^{2}\sqrt{G}dsdu{\bf =1,}\text{ }G=\det
[G_{ij}].
\end{equation}%
It is convenient to introduce a new function $\Psi ^{\prime }=\Psi
G^{1/4}$ to eliminate $\sqrt{G}$ from the area element in the
normalization condition.

After some substitution, the Schr\"{o}dinger equation is rewritten
and the new function $\Psi ^{\prime }$ is separable in $s$ and $u$,
$\Psi ^{\prime }(s,u)=\Phi (s)R(u)$, since $V(u)$ does not depend on
$s$. The eigenfunction $R(u)$ depicts the transversal motion in the
confining potential $V(u)$. For small $\varepsilon $ or deep
confining potential, the confining energy in the transversal
direction is much larger than the SO energy and the kinetic energy
in the longitudinal direction. This allows us to divide the electron
transversal motion from the longitudinal motion and the SOI. In the
limit of 1D wires, $\epsilon \rightarrow 0$, the electron will be in
the lowest transversal mode $R_{0}(u)$. Then we have an infinitely
degenerate set of states $\Psi _{n}^{\prime }(s,u)=\Phi
_{n}(s)R_{0}(u)$ where the $\Phi _{n}(s)$ denote a complete set of
spinors in the $s$ direction. Accordingly the corresponding
effective Hamiltonian can be found in a way similar to Ref. 24.

In order to obtain the exact expression of the effective
Hamiltonian, we have to calculate the lowest transversal mode for a
given confining potential. For generality and simplicity, we assume
the confining potentials
to be symmetric: for example, the harmonic potential [$V(u)=\frac{1}{2}%
Ku^{2} $], the hard-wall potential or the $\delta (u)$ potential. We
find that all these different potentials give the same results as a
result of the symmetry of the potential. The 1D effective
Hamiltonian in the presence of
SOI for a planar quantum wire is given by%
\begin{equation}
\hat{H}_{1D}=-\frac{\hbar ^{2}}{2m}\frac{\partial ^{2}}{\partial s^{2}}-%
\frac{\hbar ^{2}\kappa ^{2}}{8m}-i\alpha \lbrack \sigma _{b}\frac{\partial }{%
\partial s}-\frac{1}{2}\sigma _{t}\kappa ],
\end{equation}%
where $\sigma _{b}={\bf \hat{\sigma}}\cdot $ ${\bf b}$ and $\sigma
_{t}={\bf \hat{\sigma}}\cdot $ ${\bf t=\hat{\sigma}}\cdot
\frac{d{\bf a}(s)}{ds}$, expressed by the usual Pauli spin matrices
$\hat{\sigma}_{x,y,z}$, are the spin matrices on the normal and
tangent direction in the curvilinear coordinate respectively.\ Here
we have redefined $\alpha $ ($\alpha \rightarrow \hbar \alpha $).
Obviously, the equation gives an effective geometric potential term
$V_{eff}(s)=-\hbar ^{2}\kappa ^{2}/8m$ in agreement with the term
given by da Costa$^{[23]}$.

Specially the last term in Eq. (5)
\begin{equation}
\hat{H}_{1D,SO}=-i\alpha \lbrack \sigma _{b}\frac{\partial }{\partial s}-%
\frac{1}{2}\sigma _{t}\kappa ],
\end{equation}%
stands for the Rashba SOI in a planar curved quantum wire. It can be
easily seen that the geometry of the wire, via the curvature $\kappa
$, has influence on the SOI. Note that this 1D SOI\ Hamiltonian is
universal for planar wire with arbitrary shape. As a consequence, we
can get the
corresponding Hamiltonian for a ring which is the same as Meijer {\it et al}%
' s$^{[24]}$.

\section{QUANTUM TRANSPORT IN A 1D PLANAR CURVILINEAR QUANTUM WIRE}

The structure of the device we consider is shown schematically in
Fig. 1(b),
consisting of two straight lines conjugated with an arc of a circumference$%
^{[25,26]}$. Due to the presence of geometric potential, the curved
section can be represented as a rectangular potential well (see Fig.
1(d)) with a
width $\theta _{w}\rho $ and a depth $-\hbar ^{2}/(8m\rho ^{2})$, where $%
\rho $ is the radius of the circumference and $\theta _{w}$ is the
angle between two rectilinear parts of the wire. There exits one and
only one bound state for $\theta _{w}<\pi $ in such a
system$^{[15]}$. In order to study the problem of electron
transportation, we need to investigate the eigenstates of this
system first.

\subsection{Eigenstates and Energy Spectrum}

For convenience, the device is divided into three parts as shown in
the Fig. 1(c), and the two straight lines are assumed to be
semi-infinite. To take into account the geometrical effects, the SOI
is considered for all the three sections. Using the above 1D\
effective Hamiltonian, we can easily write the Hamiltonians for each
part. In the selective coordinate of panel
(c) in Fig. 1, they read%
\begin{align}
\hat{H}_{1}& =-\frac{\hbar ^{2}}{2m}\frac{\partial ^{2}}{\partial s^{2}}%
-i\alpha \sigma _{b_{1}}\frac{\partial }{\partial s},\text{ }s\in (-\infty ,%
\text{ }0), \\
\hat{H}_{2}& =-\frac{\hbar ^{2}}{2m\rho ^{2}}\frac{\partial
^{2}}{\partial \theta ^{2}}-\frac{\hbar ^{2}}{8m\rho
^{2}}-\frac{i\alpha }{\rho }[\sigma _{b_{2}}\frac{\partial
}{\partial \theta }-\frac{1}{2}\sigma _{t_{2}}],\text{
}\theta \in \lbrack 0,\text{ }\theta _{w}],  \nonumber \\
\hat{H}_{3}& =-\frac{\hbar ^{2}}{2m}\frac{\partial ^{2}}{\partial s^{2}}%
-i\alpha \sigma _{b_{3}}\frac{\partial }{\partial s},\text{ }s\in
(\theta _{w}\rho ,\text{ }\infty ),  \nonumber
\end{align}%
where we have introduced the polar angle $\theta $ for the second
part and
substituted $\theta $ for $s$. The spin operators are in the forms%
\begin{align}
\sigma _{b_{1}}& =-\sigma _{x},\text{ }\sigma _{b_{3}}=-\sigma
_{x}\cos
\theta _{w}-\sigma _{y}\sin \theta _{w}, \\
\sigma _{b_{2}}& =-\sigma _{x}\cos \theta -\sigma _{y}\sin \theta ,\text{ }%
\sigma _{t_{2}}=-\sigma _{x}\sin \theta +\sigma _{y}\cos \theta .
\nonumber
\end{align}

We label the eigenstates of the system $\Phi_{j;\lambda}^{k,\mu}$ where $%
j=1,2,3$ denote the corresponding region of the device respectively, $%
\lambda=\pm$ indicates the direction of motion along the wire and
$\mu=\pm$ is the spin states.

For the two parts of straight line, the energy spectrum
$E_{1,3}^{\mu }$ and unnormalized eigenstates $\Phi _{1;\lambda
}^{k,\mu }$ and $\Phi _{3;\lambda
}^{k,\mu }$ are found to be%
\begin{align}
E^{\mu }& =\frac{\hbar ^{2}k^{2}}{2m}-\mu \alpha k, \\
\Phi _{1;\lambda }^{k,\mu }& =e^{i\lambda ks}\chi _{1}^{\zeta
},\text{ }\Phi _{3;\lambda }^{k,\mu }=e^{i\lambda ks}\chi
_{3}^{\zeta },
\end{align}%
where $\zeta $ takes $+$ at $\lambda =\mu $, or else $\zeta =-$.
Here the spinors are given by
\begin{equation}
\chi _{1}^{\pm }=\left(
\begin{array}{c}
\frac{\sqrt{2}}{2} \\
\pm \frac{\sqrt{2}}{2}%
\end{array}%
\right) ,\text{ }\chi _{3}^{\pm }=\left(
\begin{array}{c}
\frac{\sqrt{2}}{2} \\
\pm \frac{\sqrt{2}}{2}e^{i\theta _{w}}%
\end{array}%
\right) .
\end{equation}%
Here we take the wave vector $k\geq 0$.

For the second part, i.e. the curved section, the corresponding
eigenfunctions and eigenvalues are obtained similarly as

\begin{equation}
\Phi _{2;\lambda }^{n,\mu }=e^{i\lambda n\theta }\chi _{2}^{\zeta },
\end{equation}%
where the orbit quantum number $n\geq 0$ and $\zeta $ is the same as
Eq. (10). Here the spinors $\chi _{2}^{\zeta }$ take the forms of

\begin{equation}
\chi _{2}^{+}=\left(
\begin{array}{c}
\sin \frac{\varphi }{2} \\
e^{i\theta }\cos \frac{\varphi }{2}%
\end{array}%
\right) ,\text{ }\chi _{2}^{-}=\left(
\begin{array}{c}
\cos \frac{\varphi }{2} \\
-e^{i\theta }\sin \frac{\varphi }{2}%
\end{array}%
\right) ,
\end{equation}%
where the angle $\varphi $, satisfying $\varphi \longrightarrow \pi
/2$ in the adiabatic limit $\rho \longrightarrow \infty $, is given
by $\tan
\varphi =-\omega /\Omega $ with $\Omega =\hbar ^{2}/(2m\rho ^{2})$ and $%
\omega =\alpha /\rho $. The associated eigenenergies read%
\begin{equation}
E_{2;\lambda }^{\mu }=\Omega \lbrack (n+\lambda \frac{1}{2})^{2}-\mu
(n+\lambda \frac{1}{2})\sqrt{1+\frac{\omega ^{2}}{\Omega ^{2}}}].
\end{equation}%
We will use these results in the following section to study the\
transport properties.

\subsection{Transmission \& Reflection Coefficients and Conductance}

We consider an electron which is transported from the 1st section to
the 3rd section through the arc to study the transport properties of
the system. In this case it is appropriate to apply a spin-dependent
version of the Griffith's boundary conditions$^{[27,28]}$ at the
intersections. This reduces the electron transport through the
device to an exactly solvable 1D scattering problem. Accordingly
there are two boundary conditions: (i) the wave functions must be
continuous, and (ii) the spin probability current density must be
conserved.

Due to the conservation of energy, the total wave functions
corresponding to the same energy in the three regions can be written
as linear combination of the corresponding eigenstates. In the
presence of Rashba coupling, the energy splitting is such that
electrons with Fermi energy $E_{F}$ have different wave numbers
depending on spin ($\mu $) and the direction of
motion ($\lambda $). The quantities $k$ and $n$ are obtained by solving $%
E_{\lambda }^{\mu }=$\ $E_{F}$ in Eq. (9) and Eq. (14) respectively.
Let the incoming electron wave function $\Phi _{in}$\ be $\cos (\phi
)\Phi
_{1;+}^{k,+}+\sin (\phi )\Phi _{1;+}^{k^{\prime },-}$, where the parameter $%
\phi $ determines the probability amplitude of each eigenstate, the
wave
functions of the device read%
\begin{align*}
\Phi _{1}& =\cos (\phi )\Phi _{1;+}^{k,+}+\sin (\phi )\Phi
_{1;+}^{k^{\prime
},-}+r_{+}\Phi _{1;-}^{k^{\prime },-}+r_{-}\Phi _{1;-}^{k,+}, \\
\Phi _{2}& =c_{1}\Phi _{2;+}^{n1,+}+c_{2}\Phi
_{2;-}^{n2,+}+d_{1}\Phi
_{2;+}^{n3,-}+d_{2}\Phi _{2;-}^{n4,-}, \\
\Phi _{3}& =t_{+}\Phi _{3;+}^{k,+}+t_{-}\Phi _{3;+}^{k^{\prime },-},
\end{align*}%
respectively. Here $r_{\pm }$ stands for the reflection coefficients while $%
t_{\pm }$ are transmission coefficients for spin polarization $\chi
^{\pm }$ respectively, and $c_{1}$, $c_{2}$, $d_{1}$, $d_{2}$ are
the corresponding coefficients of the wave function. \qquad \qquad
\qquad

Using the boundary conditions, we can determine the reflection and
transmission probabilities which are expressed as the following
forms:

\begin{eqnarray}
R_{+} &=&\left| r_{+}\right| ^{2}=\frac{4[(A^{2}-B^{2})\sin (\beta
\theta _{w})\cos \phi
]^{2}}{(A+B)^{4}+(A-B)^{4}-2(A^{2}-B^{2})^{2}\cos (2\beta
\theta _{w})}, \\
R_{-} &=&\left| r_{-}\right| ^{2}=\frac{4[(A^{2}-B^{2})\sin (\beta
\theta _{w})\sin \phi
]^{2}}{(A+B)^{4}+(A-B)^{4}-2(A^{2}-B^{2})^{2}\cos (2\beta
\theta _{w})}, \\
T_{+} &=&\left| t_{+}\right| ^{2}=\frac{8A^{2}B^{2}(\cos ^{2}\phi
+\cos \varphi \cos (\varphi -2\phi )\cos [(1+2\gamma )\theta
_{w}]+\sin ^{2}(\varphi -\phi
))}{(A+B)^{4}+(A-B)^{4}-2(A^{2}-B^{2})^{2}\cos (2\beta
\theta _{w})}, \\
T_{-} &=&\left| t_{-}\right| ^{2}=\frac{8A^{2}B^{2}(\cos
^{2}(\varphi -\phi )-\cos \varphi \cos (\varphi -2\phi )\cos
[(1+2\gamma )\theta _{w}]+\sin ^{2}\phi
)}{(A+B)^{4}+(A-B)^{4}-2(A^{2}-B^{2})^{2}\cos (2\beta \theta _{w})},
\end{eqnarray}%
satisfying the normalization condition, $R_{+}+R_{-}+T_{+}+T_{-}=1$.
Here we
have introduced $A=\sqrt{(\frac{\alpha }{\hbar })^{2}+\frac{2E_{F}}{m}}$ , $%
B=\sqrt{(\frac{\alpha }{\hbar })^{2}+\frac{2(E_{F}+V)}{m}}$ and $V=\Omega /4$%
. The parameter $\beta $ and $\gamma $ are defined by $\beta =\frac{B\hbar }{%
2\rho \Omega }$ and $\gamma
=-\frac{1}{2}+\frac{1}{2}\sqrt{1+\frac{\omega ^{2}}{\Omega ^{2}}}$,
respectively.

Finally, the conductance in mesoscopic structures can be expressed
by means of the Landauer conductance formula which in our case reads
\begin{equation}
G=G_{0}%
\mathop{\textstyle\sum}%
\limits_{\mu =\pm }T_{\mu }=G_{0}\frac{16A^{2}B^{2}}{%
(A+B)^{4}+(A-B)^{4}-2(A^{2}-B^{2})^{2}\cos (2\beta \theta _{w})},
\end{equation}%
where the conductance quantum $G_{0}=e^{2}/h$.

\subsection{Numerical Results}

For numerical calculation, we consider the curved 1D quantum
channels based
on InAs. The electrons at the Fermi level in InAs have the effective mass $%
m=0.023m_{0}$ and the Fermi energy $E_{F}=11.13$ meV. From Eqs.
(15-18) and (19), we know that the device geometry, i.e. the radius
$\rho $\ of the arc and the angle\ $\theta _{w}$, will affect the
electron transmission. By using specific fabrication method, the
scale of $\rho $\ is chosen over a wide range: $\sim 10-1000$ nm.
While $\left| \theta _{w}\right| \leq \pi $ to avoid intersection.
In addition, the SO strength $\alpha $ is assumed to be a tunable
quantity too. For example, the SOI strength $\alpha $ can be
controlled by a gate voltage with typical values in the range
$(5\sim 20)\times 10^{-12}$ eVm within an InGaAs-based 2D electron
gas$^{[29,30]}$.

First of all, we investigate the transmission of single incident\
electron with the polarized spin. The\ incoming electron state is
chosen to be $\Phi _{in}=\Phi _{1;+}^{k,+}$, i.e. $\phi =0$. So one
can find that the reflection probability $R_{-}=0$. This means that
there are no spin-flip process when the electron is reflected at the
first junction. Another choice of the reversed electron spin
incoming states will give the analogous results.

We investigate the relationship of reflection and transmission
probabilities on the radius $\rho $ shown in Fig. 2. Panel (a-c) are
for different values of SOI strength $\alpha $ and fixed $\theta
_{w}=$ $\pi /2$, while panel (d-f) are for different $\theta _{w}$
at a fixed $\alpha $. From these figures, one can see that the
electron is reflected completely at $\rho =0$. This is induced by
our model that the geometric potential becomes infinite, i.e.
$V_{eff}(s)=-\infty $, when $\theta _{w}\neq 0$ and $\rho =0$. In
fact, the quantum wire has the width, so the radius $\rho $\ can not
achieve to zero for the condition of a smooth transition. Thereby
this result may not be measurable.

We only deal with the conditions of $\rho \geq 15$ nm. Generally
speaking, the reflection probabilities are very small. As $\rho $
increases, we find that the reflection probabilities $R_{+}$
decrease to zero quickly and the transmission probabilities
$T_{+}\rightarrow 1$ and $T_{-}\rightarrow 0$. For $R_{+}$, the
profiles of curves are slightly affected by the SO coupling and
mainly determined by the angle $\theta _{w}$ in panels (a) and (d).
From Eqs. (15-18), one can see that the SO strength $\alpha $ shows
its effect through the two parameters $A$ and $B$. Considering the
ranges of all of the physical quantities, the term consisted $\alpha
$ is found to be smaller than other terms by three orders of
magnitude at least. While for $T_{+}$ and $T_{-}$, another parameter
$\gamma $ need to be considered. So both the angle $\theta _{w}$ and
the SOI can affect the profiles of the curves obviously, as seen in
panels (b-c) and (e-f).

Figure 3 shows the reflection and transmission probabilities as
function of the angle $\theta _{w}$\ for several $\rho $ and $\alpha
$. It is obvious that the profiles of the curves are symmetric on
both sides of $\theta _{w}=0 $. For $\theta _{w}=0$, we get
$R_{+}=T_{-}=0$ and $T_{+}=1$ as expected. In our model, the 2nd
section of the device vanishes and the device becomes a whole
straight line when $\theta _{w}=0$. Then the Hamiltonians
$\hat{H}_{1}$ and $\hat{H}_{3}$ in Eq. (7) take the same form, the
electron state should have no change through the process of
transmission.

In Fig. 3, the reflection and transmission probabilities are shown
to be periodic functions. From equations (15-18), the periods of the
reflection
and transmission probabilities are derived as%
\begin{eqnarray}
P_{ref} &=&\frac{\pi }{\beta }, \\
P_{tra} &=&\frac{2\pi }{1+2\gamma }.
\end{eqnarray}%
The amplitudes can also be calculated. For example, the amplitude of
$R_{+}$ is ($A^{2}-B^{2})/(A^{2}+B^{2})$. Due to the same reason as
Fig. 2, the profiles of $R_{+}$ vs. $\theta _{w}$, including the
periods and the amplitudes, are mainly determined by $\rho $.

We plot the dependences of the probabilities on the SOI $\alpha $
for
different $\theta _{w}$ and $\rho $ in Figure 4. Those oscillation vs. $%
\alpha $, influenced by the device geometry, remain for all of the
curves. As increasing the SO strength, the SOI is enhanced and can
weaken the
geometric potential. So one can find that the reflection probabilities $%
R_{+} $ tend to zero and the transmission probabilities
$T_{+}\rightarrow 1$ and $T_{-}\rightarrow 0$ with the increscent
$\alpha $. When the SOI is absent, the spin state of the outgoing
electron is mainly determined by the angle $\theta _{w}$.

Now we turn to the electron conductance. Figs. 5(a)-(b) show the
conductance
vs. $\rho $ for different set of $\theta _{w}$ and $\alpha $. As increasing $%
\rho $, the conductance increases rapidly and goes to $G_{0}$
oscillatorily. The changes are mainly effected by the angle $\theta
_{w}$. While the relationships of the conductance to the angle
$\theta _{w}$, plotted in Figs. 5(c) and 5(d). Obviously, one can
see that the conductance is
periodical and symmetrical. Finally the dependences of the conductance on $%
\alpha $ are shown in Figs. 5(e) and 5(f). The conductance shows
oscillation and goes to $G_{0}$ along with the increasing SOI
strength. As a whole, we get $G=G_{0}(1-$ $R_{+})$ using the
normalization condition. So the
conductance has the similar behavior with the reflection probability $R_{+}$%
. For further quantitative analysis, we can see that the effects of
the device geometry and the SOI are very small and the conductance
shows a periodic variation in the $10^{-4}$. So these effects on the
conductance may be difficult to be observed.

Furthermore, we consider the electron energy in a wide range, not
only localized on\ the Fermi level. Fig. 6 shows the relationship of
the conductance versus the electron energy $E$ in units of $E_{F}$
for different values of the SOI strength $\alpha $, the angle
$\theta _{w}$ and the radius $\rho $. One can see that the
conductance also has the behavior of oscillation and goes to $G_{0}$
with the increase of $E$. Especially, there are resonant
transmissions at certain energy, similar to the scattering problem
of typical square well potential.

Finally, we turn to the incoming electron with the mixed state. From
Eq. (19), One can easily see that the conductance is independent of
the parameter $\phi $. In order to study the influence on the
transmission for different incoming electron state, we calculate the
polarization ratio $\tau =(T_{+}-T_{-})/(T_{+}+T_{-})$ for the
outgoing electron. Considering Eqs.
(17) and (18), the polarization ratio $\tau $ takes the form%
\begin{equation}
\tau =\frac{T_{+}-T_{-}}{T_{+}+T_{-}}=\cos (\varphi )\cos (\varphi
-2\phi )\cos [(1+2\gamma )\theta _{w}]+\sin (\varphi )\sin (\varphi
-2\phi ).
\end{equation}%
Fig. 7 shows the dependences of $\tau $ on the parameter $\phi $ for
different values of $\rho $, $\alpha $ and $\theta _{w}$. It is
obvious that the profiles of all curves are periodical on $\phi $.
The period of the polarization ratio $\tau $\ can be found to be
$\pi $ through Eq. (22). The outgoing electron state can be spin
polarized when we choose the appropriate
parameters. Especially for the spin polarized incident electron, such as $%
\phi =0$, \ we investigate the relationship of the polarization ratio $\tau $%
\ on the device geometry and the SOI shown in Fig. 8. We find that
the spin state of the outgoing electron can also be polarized and
just rotated around
the ${\bf n}$ direction with the angle $\theta _{w}$ as the condition of $%
\cos [(1+2\gamma )\theta _{w}]=1$ is satisfied.

\section{CONCLUSIONS AND REMARKS}

In conclusion, we have derived the 1D effective Hamiltonian for a
planar curvilinear quantum wire in the presence of Rashba SOI. The
1D form of the Rashba SOI is obtained analytically. Then we consider
the electron transport in a quantum wire consisting of two straight
lines conjugated with an arc of a circumference. For the electron
transportation of spin polarized incidence, we calculate the
reflection and transmission probabilities and find that the electron
transmission is influenced by three parameters: the device geometry
($\theta _{w}$ and $\rho $) and the SOI strength $\alpha $. The
relationships between them are discussed in detail through numerical
analysis. The device geometry can influence the electron
transportation obviously, so does the SO strength $\alpha $ except
for the electron reflection. There is no spin-flip reflection for
the device. The general\ behavior of oscillating versus $\rho $,
$\theta _{w}$ or $\alpha $ are found, especially for the angle
$\theta _{w}$ the electron transportation shows the periodic
character. Using the Landauer formalism, we obtain that the
conductance varies near $G_{0}$ and goes to $G_{0}$ oscillatorily.
The relationship of the conductance vs. the electron energy is also
investigated. The electron resonant transmissions occur at some
certain energy. Finally, we study the influence of the incoming
electron state on the electron transmission, especially the spin
polarization ratio for the outgoing electron. We can achieve the
rotated spin polarized electron output under certain conditions.

There is one point that should be noted here. In our system, the SOI
is considered for the whole device for the sake of the geometrical
effects. This also makes the SOI controllable\ through the electron
field easily. On the side, the system without the SOI on the two
straight lines is also investigated and will be published later.
Even so, the study does give us one way to modulate the electron
transmission through the correlative parameters, especially the
device geometry. As an example, the device is placed on an
elastomeric substrate, then the quantum wire can deform in company
with the substrate. Thereby the wire geometry changes when the
substratum deforms; accordingly the electron transmission changes.
We may determine the deformation if we can detect the changes of the
electron transmission. Furthermore, this study also suggests that we
may construct the custom-built device through the geometry method.

We thank S. Zhao, L. Zhang, R. Liang, Y. Liu, and Z. Yao for helpful
discussions. This work is supported by NSF of China under Grant No.
10374075. E. Zhang is also\ supported by the Doctoral Foundation
Grant of Xi'an Jiaotong University (XJTU) No. DFXJTU2004-10 and by
the NSF of XJTU (Grant No. 0900-573042).


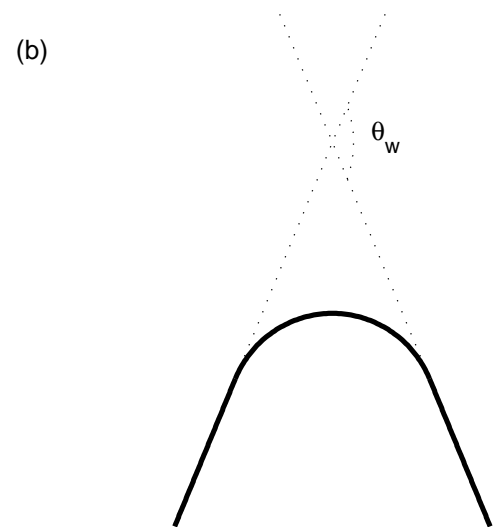
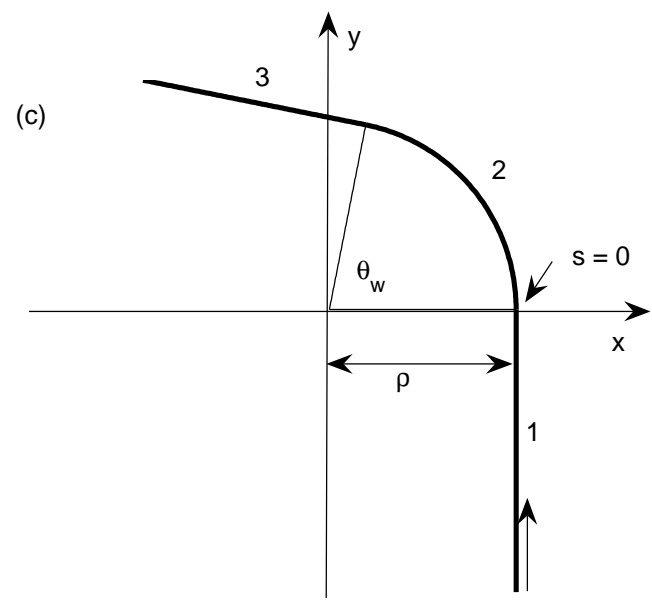
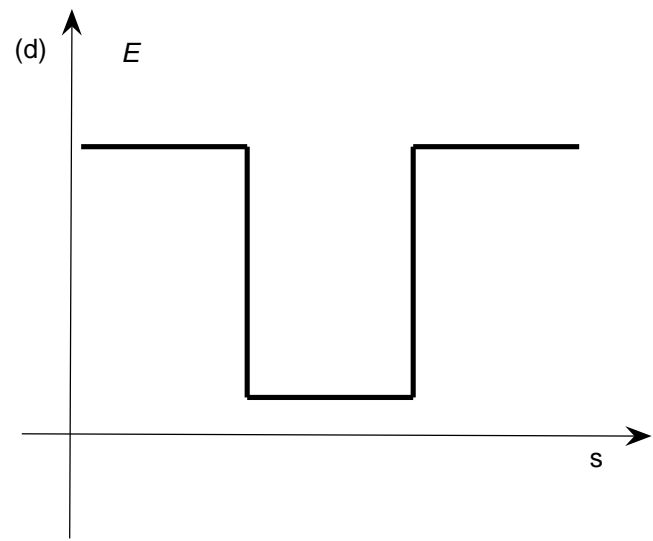
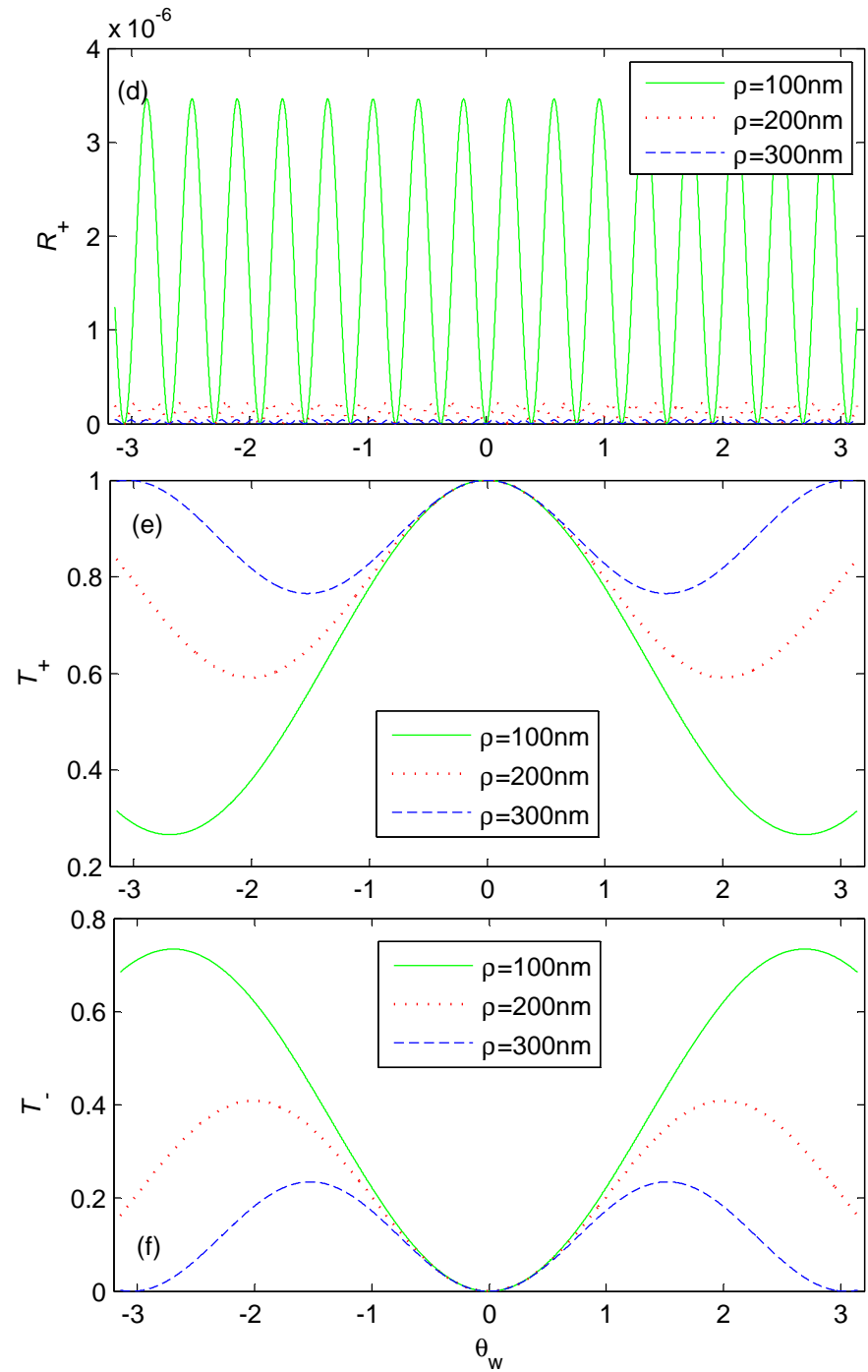
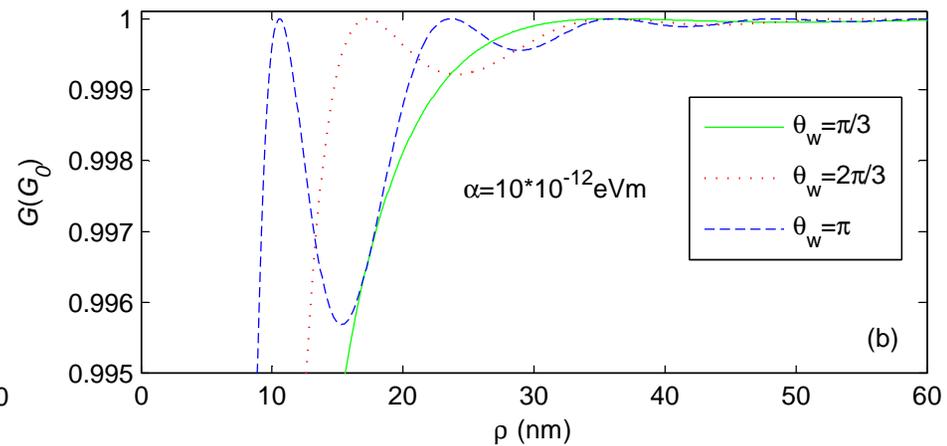
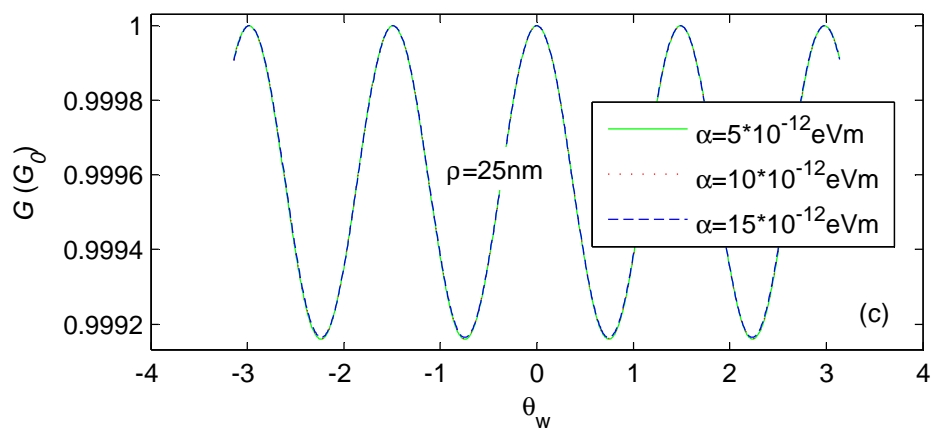
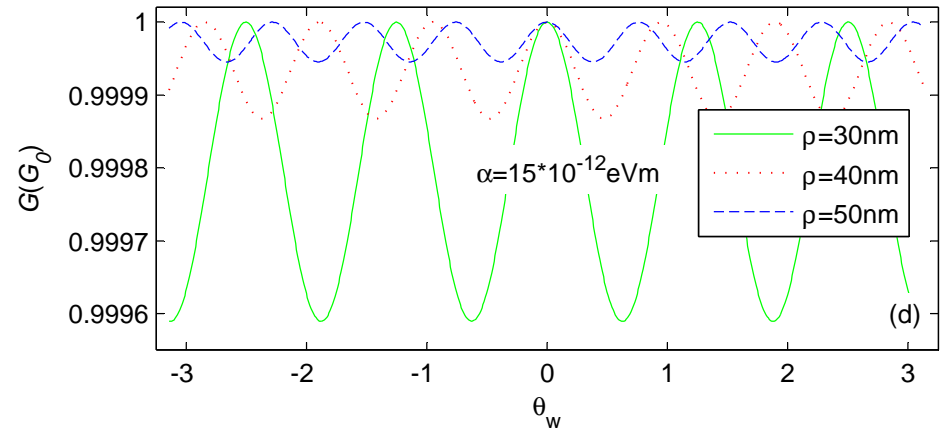
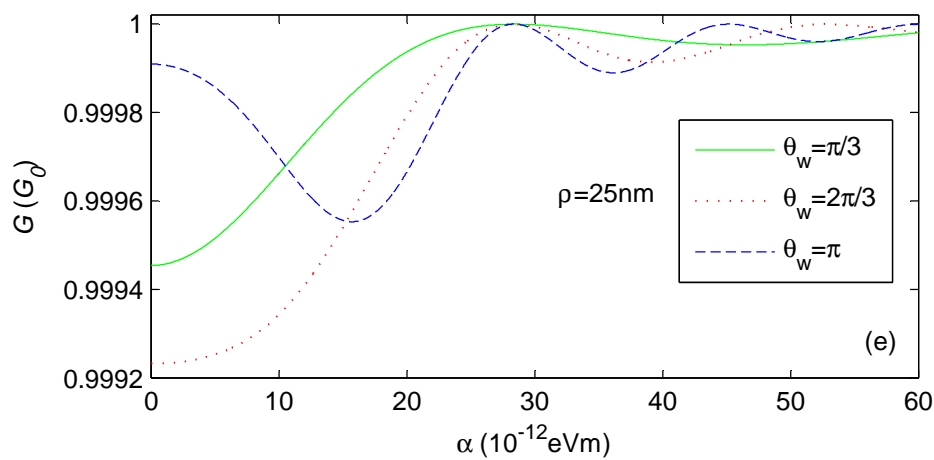
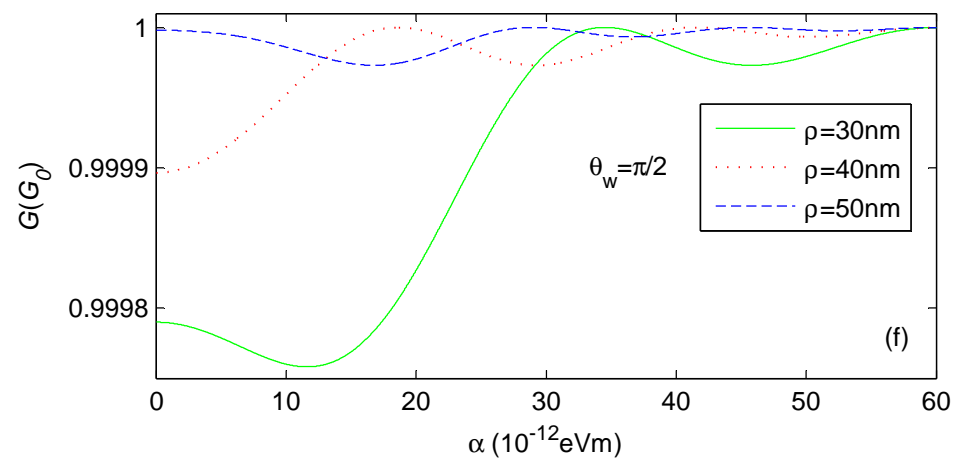
\end{document}